\documentclass[a4paper]{jpconf}
\usepackage{graphicx}

\begin{document}
\title{Top-quark mass measurements: Alternative techniques (LHC + Tevatron)}

\author{Stefanie Adomeit$^1$ on behalf of the ATLAS, CDF, CMS and D0 Collaborations}

\address{$^1$LMU M\"unchen, Fakult\"at f\"ur Physik, Schellingstra{\ss}e 4, 80799 M\"unchen, Germany}
\ead{stefanie.adomeit@physik.uni-muenchen.de}
\begin{abstract}
Measurements of the top-quark mass employing alternative techniques are presented, performed by the D0 and CDF collaborations at the Tevatron as well as the ATLAS and CMS experiments at the LHC. 
The alternative methods presented include measurements using the lifetime of $B$-hadrons, the transverse momentum of charged leptons and the endpoints of kinematic distributions in top quark anti-quark pair ($t\bar{t}$) final states. 
The extraction of the top-quark pole mass from the $t\bar{t}$ production cross-section and the normalized differential $t\bar{t}$ + 1-jet cross-section are discussed as well as the top-quark mass extraction using fixed-order 
QCD predictions at detector level. Finally, a measurement of the top-quark mass using events enhanced in single top t-channel production is presented.
\end{abstract}

\vspace{-0.4cm}
\section{Introduction}
The first world combination of top-quark mass ($m_t$) measurements has recently been performed by the CDF~\cite{cdf}, D0~\cite{d0}, ATLAS~\cite{atlas} and CMS~\cite{cms} experiments, resulting in a total uncertainty of 0.76 GeV (0.44\%)~\cite{massCombo}. All input measurements 
to the combination use standard techniques based on template, ideogram and matrix element methods which all rely on Monte-Carlo (MC) simulation to calibrate the measurement. The mass of the top quark 
implemented in MC generators, however, may differ by $\mathcal{O}\left(1\, \mathrm{GeV} \right)$ from the theoretically well-understood pole-mass~\cite{poleMC}. 
Alternative techniques can be exploited allowing for a less ambiguous theoretical interpretation of the measured top-quark mass value which becomes even more crucial now that measurements have reached sub-GeV precision. 
These alternative techniques include methods with a minimum dependence on MC simulation, for example accomplished by 
confronting the data to theory calculations. In addition, 
production modes other than top quark anti-quark pair production ($t\bar{t}$) can be studied or alternative top-quark mass estimators can be used 
to perform measurements with a different sensitivity to systematic uncertainties compared to the standard techniques.

 \section{Alternative observables}
Standard techniques to measure the top-quark mass usually make use of observables obtained via kinematic reconstruction of the $t\bar{t}$ decay (see Ref~\cite{KLFitter} for a detailed discussion on 
kinematic reconstruction techniques). Unless in-situ constraints are applied these observables 
have a strong sensitivity to uncertainties on the jet energy scale. Alternative variables such as the transverse decay length of the $B$-hadron ($L_{xy}$) or the transverse momentum of the charged 
lepton from the top quark decay show a nearly linear $m_t$-dependence. Since the reconstruction of these observables is based on the tracking systems and the electromagnetic calorimeter (electron $p_T$), analyses based on $L_{xy}$ or the 
lepton $p_T$ are less prone to variations in the jet energy scale, 
however, at the cost of an 
enhanced statistical uncertainty w.r.t. the standard approach and a large sensitivity to the modelling of the boost of the top quark in MC simulation.

CDF has used both the lepton p$_T$ and $B$-hadron lifetime separately and in a simultaneous fit to measure the top-quark mass in a dataset corresponding to L = 1.9 fb$^{-1}$ in $\ell$+jets final states~\cite{CDFLxy}.
The simultaneous fit to the median $L_{xy}$ and lepton $p_T$ distributions gives $m_t = 170.7 \pm 6.3\, (\mathrm{stat.}) \pm 2.6\, (\mathrm{syst.})\, \mathrm{GeV}$. Performing a simultaneous fit results in a reduced statistical uncertainty w.r.t. the separate measurements 
as the momentum of the leptons is mostly uncorrelated with that of the $b$-quarks. The calorimeter JES uncertainty of the measurement yields 0.3 GeV and is smaller compared to the measurements based on standard mass estimators. 

CMS has performed a top-quark mass measurement based on the $L_{xy}$ variable 
in a dataset corresponding to L = 19.3-19.6 fb$^{-1}$ and recorded at $\sqrt{s}$ = 8 TeV~\cite{CMSLxy}, using $e$/$\mu$+jets and $e\mu$ final states. Combining the three channels the mass of the top-quark is measured to be 
$m_t = 173.5 \pm 1.5\, (\mathrm{stat.}) \pm 1.3\, (\mathrm{syst.}) \pm 2.6\, (p_T^{\mathrm{top}})\, \mathrm{GeV}$, with the dominant source of systematic uncertainty stemming from the modelling of the top-$p_T$.
 
 \section{Kinematic endpoints}
 \begin{figure}[h]
 \begin{minipage}{10.2cm}
\includegraphics[height=3.6cm]{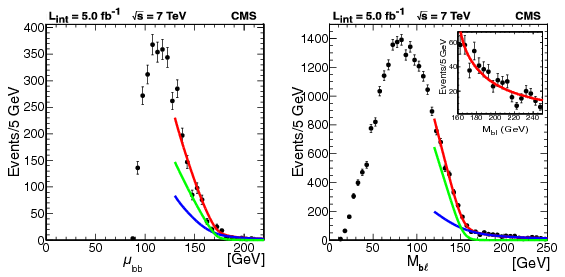}
\end{minipage}
\begin{minipage}{5.1cm}
\includegraphics[height=3cm]{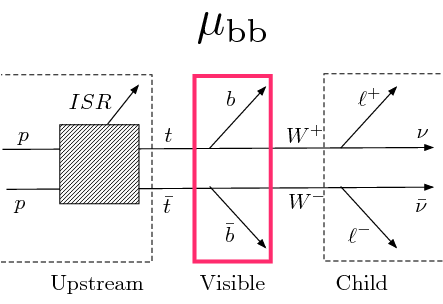}
\end{minipage}
\caption{\label{endpoints} Doubly-constrained fit to the endpoints of the $\mu_{bb}$ and $M_{b\ell}$ distributions.
The red line is the full fit while the blue and green curves are for the background and signal shapes, respectively.
The $\mu_{bb}$ variable 
uses the $b$ jets, and treats the $W$ bosons as lost
child particles~\cite{CMSKinEnd}.
}
\end{figure}

As an alternative approach the top-quark mass can be measured from the endpoints of kinematic distributions. This method was originally developed 
to determine unknown masses in decay chains in beyond standard model physics scenarios where undetected particles lead to underconstrained kinematics.
CMS has applied the method to $t\bar{t}$ final states in the dilepton channel~\cite{CMSKinEnd} which give rise to two neutrinos escaping detection per $t\bar{t}$ decay. 
The measurement is based on the transverse mass $M_{T2}$, defined as the minimum parent mass consistent with the observed kinematics:
\begin{equation}
\label{endpointF}
M_{T2}\equiv \min \limits_{\mathbf{p_{T}^{\nu_a}}+\mathbf{p_{T}^{\nu_b}}=\mathbf{p_{T}^{miss}}}\{\max(m_T^a,m_T^b)\}.
\end{equation}
Since the endpoint of the $M_{T2}$ distribution is sensitive to the parent mass it can be employed as a mass estimator. To solve the dileptonic event kinematics, the $t\bar{t}$ decay is partitioned to create three variables --
including two $M_{T2}$ subsystems and the invariant mass of the $b$-jet + charged lepton system ($M_{b\ell}$) (see Fig.~\ref{endpoints}). 
To remove the dependence of the 
$M_{T2}$ variable on the transverse momentum of the upstream top quarks, $M_{T2\perp}$ is introduced, which uses only momentum components 
perpendicular to the boost direction of the top quark.
Constraining the $\nu$ and $W$-boson masses to $m_{\nu}$ = 0 GeV and $m_{W}$ = 80.4 GeV respectively, the top-quark-mass is measured to be $m_t = 173.9 \pm 0.9\, (\mathrm{stat.}) ^{+1.7}_{-2.2}\, (\mathrm{syst.})\, \mathrm{GeV}$,
using a dataset corresponding to L = 5.0 fb$^{-1}$ and recorded at $\sqrt{s}$ = 7 TeV. The measurement is based on analytic endpoint formulas and thus provides 
a model-independent technique to determine the mass of the top-quark. Performing a simultaneous fit to the endpoints of the 3 distributions, the top-quark, $W$-boson and 
$\nu$ masses can be determined, allowing for a test of mass determination methods that may be used in beyond standard model physics scenarios.

\section{Top-quark mass from the $t\bar{t}$ cross-section}
The theoretical dependence of the $t\bar{t}$ cross-section on $m_t$ can be exploited to extract the mass of the top-quark from measurements of $\sigma_{t\bar{t}}$. Comparing the measured cross-section to the theory 
calculation, the top-quark mass can be measured unambiguously in the renormalization scheme used in the $\sigma_{t\bar{t}}(m_t)$ calculation. The criteria applied to select the 
$t\bar{t}$ sample for the $\sigma_{t\bar{t}}$ measurement, however, introduce a residual dependence of the acceptance -- and, hence, of the measured cross-section -- on the top-quark mass implemented in MC simulation. 
Usually $m_t^{MC}=m_t^{pole}$ is assumed to parameterize the $m_t$-dependence of the measured $t\bar{t}$ cross-section in these measurements.

The top-quark mass has been measured in both the pole-mass and $\overline{MS}$ scheme at approximate NNLO by D0, using $\ell$ + jets final states~\cite{xSecD0} in a dataset corresponding to L = 5.3 fb$^{-1}$. 
Assuming $m_t^{MC}=m_t^{pole}$ for the experimental 
$m_t$-dependence of the measured $t\bar{t}$ cross-section yields a pole ($\overline{MS}$) mass of $m_{t}^{pole} = 167.5^{+5.2}_{-4.7}$ GeV ($m_{t}^{\overline{MS}} = 160.0^{+4.8}_{-4.2}$ GeV) while the 
alternative scenario with $m_t^{MC}=m_t^{\overline{MS}}$ has been checked and found to result in a shift of 2.5 GeV (2.6 GeV) of the extracted pole ($\overline{MS}$) mass. CMS and ATLAS have measured the 
top-quark pole mass in dileptonic $t\bar{t}$ final states at NNLO+NNLL accuracy~\cite{xSecCMS,xSecATLAS}. Using data recorded at $\sqrt{s}$ = 7 TeV (L = 2.3 fb$^{-1}$), CMS measures a top-quark pole mass of 
$m_t^{\mathrm{pole}}=176.7 ^{+3.0}_{-2.8}\, \mathrm{GeV}$ while ATLAS has performed 
a combined measurement using datasets recorded at $\sqrt{s}$ = 7 TeV (L = 4.6 fb$^{-1}$) and $\sqrt{s}$ = 8 TeV (L = 20.3 fb$^{-1}$), obtaining $m_t^{\mathrm{pole}}=172.9 ^{+2.5} _{-2.6}\, \mathrm{GeV}$. For the latter measurement 
the experimental $t\bar{t}$ cross-section shows only a marginal dependence of $d\sigma/dm_t = -0.28 \pm 0.03/\mathrm{GeV}$ on $m_t^{MC}$ (see Fig.~\ref{crossSection}).

\section{Top-quark mass from the $t\bar{t}$+1 jet cross-section}
\begin{figure}[t]
 \begin{minipage}[t]{7.6cm}
\includegraphics[height=4.3cm]{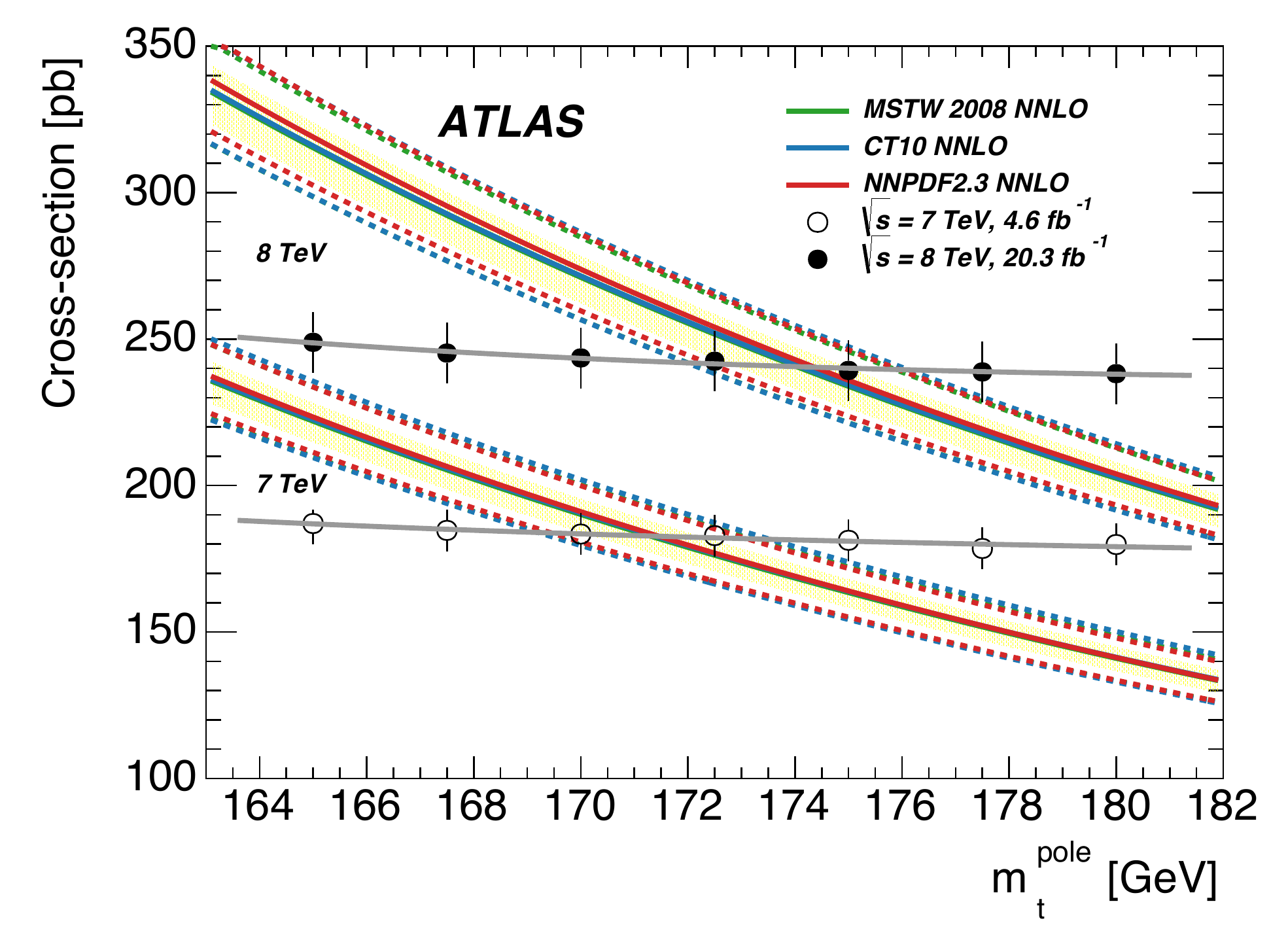}
\caption{\label{crossSection} Predicted NNLO+NNLL $\sigma(t\bar{t})$ 
at $\sqrt{s}$ = 7 TeV and $\sqrt{s}$ = 8 TeV as a function of $m_t^{pole}$, showing 
the central values and total uncertainty bands with several PDF sets. The yellow band shows the QCD scale uncertainty. The measurements of $\sigma(t\bar{t})$ are also shown, with their 
dependence on the assumed value of $m_t$ through acceptance and background corrections~\cite{xSecATLAS}.}
\end{minipage} \hspace{0.2cm}
\begin{minipage}{7.4cm}
\includegraphics[height=4.3cm]{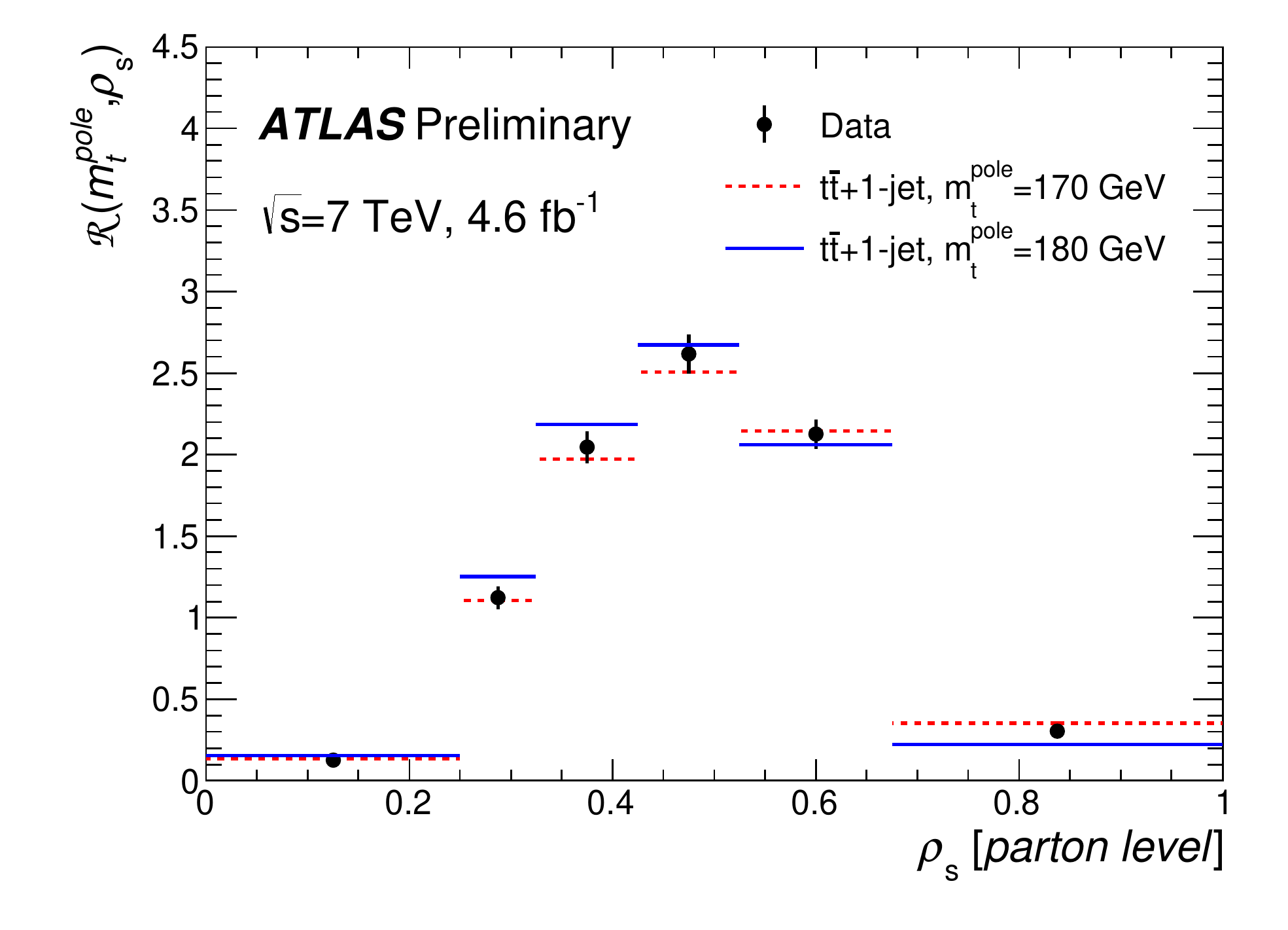}
\caption{\label{diffCrossSection} $\mathcal{R}(m_t^{pole},\rho_{s})$ 
distribution at parton level corrected for detector and hadronization effects after the background subtraction. The red-dotted (blue-continuous) lines 
correspond to the $t\bar{t}$+1-jet @ NLO+PS calculation using $m_t^{pole}$ = 170 GeV (180 GeV). The black points correspond to the data~\cite{xSecDiffATLAS}.}
\end{minipage}
\end{figure}

A novel technique to determine the top-quark pole mass has been introduced in~\cite{xSecDiffTh} which is based on the normalized $t\bar{t}$ + 1 jet cross section, differential in the invariant mass of the 
final state jets:
\begin{equation}
  \mathcal{R}(m_t^{pole},\rho_{s})=\frac{1}{\sigma_{t\bar{t}+\mathrm{1-jet}}}\frac{d\sigma_{t\bar{t}+\mathrm{1-jet}}}{d\rho_{s}}(m_t^{pole},\rho_{s}),
 \end{equation}
 with $\rho_{s}=\frac{2m_{0}}{\sqrt{s_{t\bar{t}j}}}$ and $m_0=170\, \mathrm{GeV}$. This method has been applied by ATLAS, using $\ell$+jets final states 
 in a dataset corresponding to L = 4.6 fb$^{-1}$, recorded at $\sqrt{s}$ = 7 TeV~\cite{xSecDiffATLAS}. The $t\bar{t}$ + 1 jet system is reconstructed 
 via a kinematic fit, where events are kept if the highest transverse momentum jet not assigned to the $t\bar{t}$ decay fulfills p$_T > 50\, \mathrm{GeV}$. 
 To extract the top-quark pole mass the $\mathcal{R}(m_t^{pole},\rho_{s})$ distribution in data is unfolded to parton level and compared to the theoretical prediction at NLO+PS accuracy (see Fig.~\ref{diffCrossSection}), yielding 
 $m_t^{\mathrm{pole}}=173.7 \pm 1.5\, (\mathrm{stat.}) \pm 1.4 \, (\mathrm{syst.})\, ^{+1.0}_{-0.5}\, (\mathrm{theo.})\, \mathrm{GeV}$. The theoretical uncertainties on the 
 NLO+PS prediction -- including PDF, renormalization and factorization scale uncertainties -- are largely reduced due to the usage of the normalized cross-section. The systematic uncertainties on the 
 measured $\mathcal{R}(m_t^{pole},\rho_{s})$ distribution are dominated by uncertainties on the jet energy scale.
 
 \section{Using fixed order calculations and forward folding}
\begin{figure}[h]
 \begin{minipage}{7.0cm}
\includegraphics[height=4.3cm]{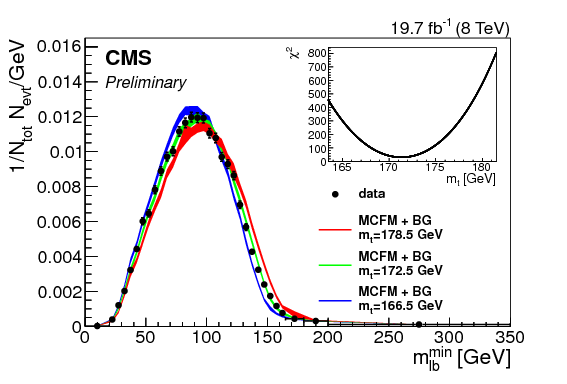}
\end{minipage} \hspace{0.1cm}
\begin{minipage}{8.4cm}
\caption{\label{mlbFig} Normalized event yield for $t\bar{t}$ production at $\sqrt{s}$ = 8 TeV, presented as a function of $m_{\ell b}$. 
The observed yields are shown by closed symbols, with error bars representing statistical uncertainty. For illustration, the MCFM (NLO in production) 
predictions at $m_t$ = 178.5 GeV (red band), $m_t$ = 172.5 GeV (green band) and $m_t$ = 166.5 GeV (blue band) are shown~\cite{CMSDil}. 
}
\end{minipage}
\end{figure}

 Direct top-quark mass measurements make use of MC samples -- generated at different top-quark input masses -- to parameterize the shapes of the mass estimator, usually invariant mass distributions,
 as a function of $m_t$. In addition to this approach, CMS has used fixed-order NLO calculations -- which imply an unambiguously defined 
 pole mass -- to extract the mass dependence of the invariant mass of the $b$-jet + lepton system ($m_{\ell b}$) in dileptonically decaying $t\bar{t}$ pairs~\cite{CMSDil}. To measure the top-quark mass, 
 the shape of the $m_{\ell b}$ distribution in data is compared to the fixed-order calculation MCFM~\cite{MCFM2,MCFM3} in the visible phase-space (see Fig.~\ref{mlbFig}). 
 Since the analysis is performed at reconstruction level the predicted $m_{\ell b}$ distribution $\vec{x}_{pred}$ is folded to the reconstruction level $\vec{x}_{reco}$
 \begin{equation}
 \vec{x}_{reco} =   \mathcal{L} \ \mathcal{M}^{resp}\vec{x}_{pred},
 \end{equation}
 where a dedicated response matrix $\mathcal{M}_{ij}^{resp}$ is calculated for each systematic variation and seven values of $m_t$ based on MC simulation (MadGraph~\cite{MadGraph} + Pythia~\cite{Pythia}). 
 The several response matrices as well as the measured $m_{\ell b}$ distribution will be made publicly available.
 Using a dataset corresponding to an integrated luminosity of L = 19.7 fb$^{-1}$ and recorded at $\sqrt{s}$ = 8 TeV the shape fit of the NLO MCFM prediction to the measurement yields 
 $m_t = 171.4 ^{+1.0}_{-1.1}$ GeV. The result is consistent with the top-quark mass obtained when using MC simulation matched to parton shower (MadGraph+Pythia) instead of the fixed-order MCFM calculation, 
 which gives $m_t = 172.3 \pm 1.3$ GeV.

\section{Top-quark mass measurement in single top $t$-channel enhanced events}
For the first time the top-quark mass has been measured in topologies enhanced with single top quarks produced in the $t$-channel by ATLAS~\cite{singleTopMass}. Due to their electroweak production mode, single top final states show different sensitivity to signal modelling uncertainties compared to $t\bar{t}$ production,
such as the modelling of color flow. The measurement uses a dataset 
corresponding to an integrated luminosity of L = 20.3 fb$^{-1}$ and recorded at $\sqrt{s} = 8\, \mathrm{TeV}$. Signal events are selected using a neural network, where the training is performed only against non top-quark backgrounds to enhance the statistical sensitivity 
of the measurement. The data sample used for the measurement of the top-quark mass comprises $\approx$ 50\% single top $t$-channel, $\approx$ 23\% $t\bar{t}$ and $\approx$ 27\% non-top background events. A template method 
is applied to measure the top-quark mass, where final states with $t\bar{t}$ and all single top channels are treated as signal. The top-quark mass is measured to be $m_t = 172.2 \pm 0.7\, (\mathrm{stat.}) \pm 1.9\, (\mathrm{syst.})\, \mathrm{GeV}$, with 
the dominant source of systematic uncertainty stemming from the jet energy scale.

\section{Conclusions}
Given the sub-GeV level precision reached by the direct top-quark mass measurements, exploring and developing alternative methods has become an important task to further improve 
our understanding of the systematic uncertainties and the interpretation of the measured top-quark mass value.
These alternative methods include techniques which are largely independent of MC simulation as well as measurements using alternative observables and final states which are sensitive 
to different systematic uncertainties. Though most of the alternative measurements of the top-quark mass are not yet competitive with the standard approaches, 
their precision is ever increasing. The higher data statistics provided in Run 2 might allow for a 
further improvements of the alternative techniques and the realization of new approaches, e.g. as proposed in~\cite{JPsi} and investigated in~\cite{CMSJPsi}.
\newline
\textbf{Acknowledgments} 
\newline
The author would like to thank Benjamin Stieger for his help on the CMS analyses.

\end{document}